\documentclass[conference]{IEEEtran}
\IEEEoverridecommandlockouts
\usepackage{cite}
\usepackage{amsmath,amssymb,amsfonts}
\usepackage{algorithmic}
\usepackage{booktabs}
\usepackage{arydshln}
\usepackage{graphicx}
\usepackage{textcomp}
\usepackage{xcolor}
\usepackage{xspace}
\usepackage{soul}
\usepackage{cleveref}
\crefname{section}{Sec.}{Secs.}
\Crefname{section}{Section}{Sections}
\Crefname{table}{Table}{Tables}
\crefname{table}{Tab.}{Tabs.}

\usepackage[acronym, shortcuts, nohypertypes={acronym}]{glossaries}
\newacronym{CCC}{CCC}{concordance correlation coefficient}
\newacronym{CNN}{CNN}{convolutional neural network}
\newacronym{EXVO}{\textsc{ExVo}}{ICML Expressive Vocalisations Workshop \& Competition 2022}
\newacronym{ACII}{\textsc{A-VB}}{ACII Affective Vocal Burst Workshop \& Challenge 2022}
\newacronym{HVB}{\textsc{Hume-VB}}{The Hume Vocal Burst Competition Dataset}
\newacronym{FFNN}{FFNN}{feed-forward neural network}
\newacronym{SER}{SER}{speech emotion recognition}
\newacronym{SGD}{SGD}{stochastic gradient descent}

\newcommand{\etyou}{\textsc{End2You}}
\newcommand{\wav}{\textsc{wav2vec2.0}\xspace}
\newcommand{\avbhigh}{\textsc{A-VB-High}\xspace}
\newcommand{\avbtwo}{\textsc{A-VB-Two}\xspace}
\newcommand{\avbculture}{\textsc{A-VB-Culture}\xspace}
\newcommand{\avbtype}{\textsc{A-VB-Type}\xspace}

\def\BibTeX{{\rm B\kern-.05em{\sc i\kern-.025em b}\kern-.08em
    T\kern-.1667em\lower.7ex\hbox{E}\kern-.125emX}}

\makeatletter
\newcommand{\linebreakand}{%
  \end{@IEEEauthorhalign}
  \hfill\mbox{}\par
  \mbox{}\hfill\begin{@IEEEauthorhalign}
}
\makeatother

\makeatletter
\DeclareRobustCommand\onedot{\futurelet\@let@token\@onedot}
\def\@onedot{\ifx\@let@token.\else.\null\fi\xspace}
\def\eg{\emph{e.g}\onedot}

\def\etal{\emph{et al}\onedot}
\makeatother

\begin{document}

\title{Self-Supervised Attention Networks and Uncertainty Loss Weighting for Multi-Task Emotion Recognition on Vocal Bursts\\

\thanks{This research received funding from the BMW Group.}
}

\author{
\IEEEauthorblockN{1\textsuperscript{st} Vincent Karas}
\IEEEauthorblockA{\textit{Chair EIHW} \\
\textit{University of Augsburg / BMW Group}\\
Augsburg, Germany \\
vincent.karas@uni-a.de}
\and
\IEEEauthorblockN{2\textsuperscript{nd} Andreas Triantafyllopoulos}
\IEEEauthorblockA{\textit{Chair EIHW} \\
\textit{University of Augsburg}\\
Augsburg, Germany \\
andreas.triantafyllopoulos@uni-a.de}
\and
\IEEEauthorblockN{3\textsuperscript{rd} Meishu Song}
\IEEEauthorblockA{\textit{Chair EIHW} \\
\textit{University of Augsburg / University of Tokyo}\\
Augsburg, Germany / Tokyo, Japan \\
meishu.song@uni-a.de}
\linebreakand
\IEEEauthorblockN{4\textsuperscript{th} Björn W. Schuller}
\IEEEauthorblockA{\textit{Chair EIHW \& ZIG / GLAM} \\
\textit{Unversity of Augsburg / Imperial College London}\\
Augsburg, Germany / London, UK\\
bjoern.schuller@uni-a.de}
}

\maketitle

\begin{abstract}

Vocal bursts play an important role in communicating affect, making them valuable for improving speech emotion recognition. Here, we present our approach for classifying vocal bursts and predicting their emotional significance in the ACII Affective Vocal Burst Workshop \& Challenge 2022 (A-VB). We use a large self-supervised audio model as shared feature extractor and compare multiple architectures built on classifier chains and attention networks, combined with uncertainty loss weighting strategies. Our approach surpasses the challenge baseline by a wide margin on all four tasks.



\end{abstract}

\begin{IEEEkeywords}
multi-task learning, wav2vec2, uncertainty loss, classifier chain, task attention network
\end{IEEEkeywords}

\section{Introduction}


Vocal bursts, such as laughs, sobs, or sighs, are vital indicators of affect, oftentimes more informative than prosody for the recognition of emotions~\cite{Hawk09-WORTH, Sauter13-CRE}.
Recent work attempts to shed more light on how much emotional information, and of what kind, humans are capable of expressing and comprehending based on vocal bursts~\cite{Cordaro16-VCE, Cowen19-BHV}.
It has shown that emotional (or affective) vocal bursts carry information for over ten emotional dimensions that can be reliably understood across different cultures.
This lays the theoretical foundation for using this information to more robustly and more holistically understand emotional reactions to external stimuli, and is gaining steam as a novel research direction in the crucifix of machine learning and affective computing.

This is being increasingly utilised by approaches seeking to improve \ac{SER} performance that work by identifying and analysing vocal bursts separately from speech signals~\cite{Huang19-NVS, Hsu21-NVA}.
The recent \ac{EXVO}~\cite{Baird22-EXVO} and the ongoing \ac{ACII}~\cite{Baird22-ACII} present a new front for further research in the study of emotional bursts, both utilising the \ac{HVB}~\cite{Cowen2022HumeVB}.
Recent advances in \ac{SER} research have shown that the use of large, transformer-based models has proven a critical asset in improving performance~\cite{Pepino2021a,Wagner22-Dawn}.
Such models have proven equally effective in the modelling of emotion from vocal bursts~\cite{Xin22-EES}.
This exceeded benefits obtained by pre-training smaller models on in-domain data~\cite{Jing22-RRT}, as well as those obtained by personalisation methods~\cite{Triantafyllopoulos22-ESR}, showing that larger models trained on bigger data can learn better and more generalisable representations which is critical for obtaining good results on \ac{HVB} data.

Further benefits have also been obtained through the use of multi-tasking, which is typically used to improve \ac{SER} performance~\cite{Wagner22-Dawn}.
In the case of \ac{EXVO}, this was achieved by combining affect with country-of-origin and age~\cite{Song22-DRU, Atmaja22-JPE}.
The exploitation of such methods is expected to be highly beneficial for several \ac{ACII} tasks as well, as emotion recognition tasks are heavily interdependent~\cite{Xin22-EES}.
In particular, two methods stand out from previous work on \ac{EXVO}.
Classifier chains~\cite{Read11-CC} were used by Xin et al.~\cite{Xin22-EES} to improve performance for few-shot, personalised emotion recognition.
This model predicts each task sequentially, with the output of earlier tasks propagated to later task layers to improve performance.
Song et al.~\cite{Song22-DRU} instead used an uncertainty weighting loss to dynamically balance the contribution of the age, country and emotion 
losses during training, which was shown to improve performance over a standard averaging of the losses.
These methods show how the emotional tasks contained in the \ac{HVB} dataset can be properly combined. 

In the present work, we draw on those recent advances by utilising the learnt representations of a pre-trained \wav~\cite{Baevski20-W2V} in conjunction with classifier chains~\cite{Read11-CC} and uncertainty weighting losses~\cite{Song22-DRU} to perform multi-task learning on all four \ac{ACII} tasks: \avbhigh, where ten emotional dimensions are predicted on a continuous scale; \avbtwo, where the two affect dimensions of arousal and valence are predicted; \avbculture, where culture-specific predictions for the ten high-level dimensions are expected; and \avbtype, where seven types of bursts need to be classified.
Our comprehensive evaluation over different experimental configurations better elucidates the strengths and weaknesses of the proposed methods for the analysis of emotional vocal bursts. We make our code available on Github\footnote{https://github.com/VincentKaras/a-vb-emotions}.

The rest of this paper is structured as follows:  We describe the models and loss weighting strategies in \cref{sec:meth}. Our experimental settings and results are presented in \cref{sec:exp}. The results are discussed in \cref{sec:discussion}. Finally, \cref{sec:conclusion} summarises the paper.




\section{Methodology}\label{sec:meth}

\begin{figure*}[ht]
\centering
\includegraphics[width=0.9\textwidth]{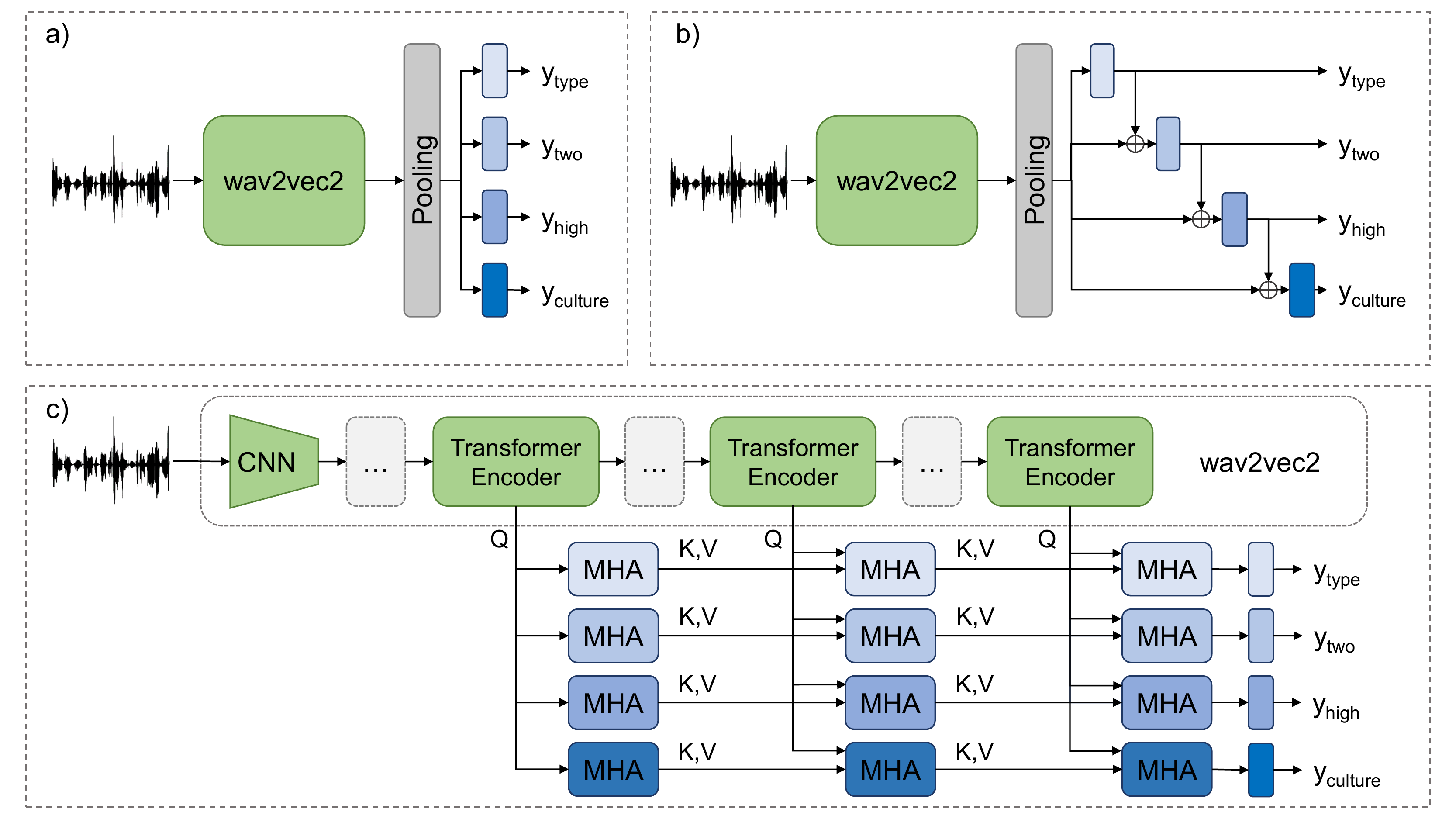}
\caption{Model architectures. a) basic mtl, b) classifier chain, c) branching multi-head attention model.}
\label{fig:models}
\end{figure*}

Our method is based on using large models pre-trained on speech with self-supervised learning (SSL). Specifically, we make use of wav2vec2 \cite{Baevski20-W2V}, which processes raw audio and combines a CNN with a stack of transformer encoder layers. It is trained by masking the input sequence to the transformer and solving a contrastive learning task over quantised latent representations.  
There is a wide selection of wav2vec2-like models available, which differ in the architecture and the corpus used for fine-tuning, \eg Librispeech. For this paper, we use the wav2vec2-base architecture\footnote{https://huggingface.co/facebook/wav2vec2-base}, without additional training on a speech corpus. This choice is motivated by the competition dataset, as there is likely no significant benefit for vocal bursts from fine-tuning towards speech. Wav2vec2-base has 12 transformer layers, which yield representations of size 768. Its CNN features have a size of 512. 
With wav2vec2-base as backbone, we construct three architectures, which are described in the following sections and illustrated in \cref{fig:models}. 

\subsection{Vanilla Model}\label{meth:base}
For our basic multitask model, we process the features in parallel networks to predict the tasks. The networks are shallow, with each having one hidden fully connected (FC) layer and one output layer.  

\subsection{Classifier Chain Model}\label{meth:chain}

This model uses a chain of classifiers, similar to the one proposed in \cite{Xin22-EES}. We use the features returned by \wav as the starting point for the first task, then concatenate them with the predictions for each task as input to the next task network. At training time, we use the ground truth as input instead of the predictions to avoid confusing the model with inaccurate guesses at the start of the training.
The order of the four tasks is chosen based on the assumption that the model will benefit by learning from low to high complexity. Thus, we begin with \avbtype, which should be easier than recognising emotional content, proceed with \avbtwo as a general representation of affective state, then continue with predicting specific emotions in \avbhigh, and end with culture specific \avbculture.


\subsection{Branching Multi-Head Attention Model}

For this model, the hidden activations of the transformer block of \wav are used as inputs to the task networks. This is based on the assumption that for each task, activations at different layers may contain useful information, which was also used for the CNN model in \cite{Song22-DRU}. However, instead of separate networks branching off individual layers in the backbone net, we use attention to combine the features from multiple layers. A similar approach was proposed in \cite{Liu2019b} with a multi-task attention network (MTAN). For each task, we construct a network of multi-head attention (MHA) layers, where the outputs of each MHA layer are used as key, and value pairs for the next block, while the hidden states of \wav form the queries.



\subsection{Losses and Metrics}

For \avbtype, categorical cross entropy is used as the loss function, and unweighted average recall (UAR) as metric. The emotion regression tasks use the concordance correlation coefficient (CCC) \cite{Lin1989} and Pearson's correlation coefficient $\rho$ as metrics. They are defined in \cref{eq:ccc} and \cref{eq:pearson}, respectively. We also use CCC as our objective function in the form $1-CCC$.

\begin{equation}
\begin{aligned}
    & CCC(x, y) = \frac{2 * cov(x,y)}{\sigma_{x}^2 + \sigma_{y}^2 + \left(\mu_x - \mu_y\right)^2}, \\
    & \text{where}\, cov(x,y) = \sum \left(x - \mu_x\right) \left(y - \mu_y\right).
\end{aligned}
\label{eq:ccc}
\end{equation}

\begin{equation}
    \begin{aligned}
        & \rho(x, y) = \frac{cox(x,y)}{\sigma_{x}\sigma{y}}. \\
    \end{aligned}
    \label{eq:pearson}
\end{equation}

While multi-task learning can achieve superior results by exploiting relationships between the labels, there is a challenge in guiding the training such that all tasks achieve good performance, instead of just the easier ones \cite{Liu2019b}. Thus, the losses need to be balanced, but choosing weights for a large number of tasks requires extensive tuning. Instead, we opt for three adaptive loss weighting strategies, with fixed equal weights as comparison.

\subsubsection{Dynamic Weight Averaging}

Dynamic Weight Averaging (DWA) \cite{liu2021towards} adapts the task weights over time based on the rate of change of the loss in previous steps. The weight is defined in \cref{eq:dwa}: 

\begin{equation}
    \begin{aligned}
        \lambda_k(t) =  \mathcal{K} \frac{ exp\left(\frac{\mathcal{L}_k(t-1)}{\mathcal{L}_k(t-2)}/\mathcal{T}\right)}{\sum_{K}exp\left(\frac{\mathcal{L}_k(t-1)}{\mathcal{L}_k(t-2)}/\mathcal{T}\right)},
    \end{aligned}
    \label{eq:dwa}
\end{equation}

which is essentially a temperature softmax function, with increasing values of the temperature $\mathcal{T}$ smoothing the weights. The factor $\mathcal{K}$ scales the weights such that they sum to the number of tasks $K$. The total loss then becomes a weighted sum as in \cref{eq:dwaloss}:

\begin{equation}
    \begin{aligned}
        \mathcal{L}_{dwa} = \sum_{K} \lambda_k(t) \mathcal{L}_k.
    \end{aligned}
    \label{eq:dwaloss}
\end{equation}

\subsubsection{Restrained Revised Uncertainty Loss}

The restrained revised uncertainty loss (RRUW) was proposed by Song \etal \cite{Song22-DRU} as a modification to \cite{liebel2018auxiliary}. It uses trainable parameters $\alpha$ to scale the task weights, which are constrained to avoid trivial solutions. The RRUW is defined in \cref{eq:rruwloss}

\begin{equation}
    \begin{aligned}
        \mathcal{L}\left(w,\alpha\right) = &\sum_K \frac{1}{\alpha_k^2} \mathcal{L}_k(w) + \sum_K \log\left(1 + \log\alpha_k^2\right) + \\
        &| \varphi - \sum_K\left( | \log\alpha_k | \right)|,
    \end{aligned}
    \label{eq:rruwloss}
\end{equation}

where $\alpha_{i}$ are the trainable weights, and $\varphi$ is a positive value that the sum of weights is driven towards.

\subsubsection{Dynamic Restrained Uncertainty Weighting}

Dynamic Restrained Uncertainty Weighting (DRUW) combines the dynamic and uncertainty weights of DWA and RRUW. Thus, the final loss becomes:

\begin{equation}
    \begin{aligned}
         \mathcal{L}\left(w,\alpha\right) = &\sum_K \left( \frac{1}{\alpha_k^2} + \lambda_k(t) \right) \mathcal{L}_k(w) + \\
         &\sum_K \log\left(1 + \log\alpha_k^2\right) + | \varphi - \sum_K\left( | \log\alpha_k | \right)|.
    \end{aligned}
    \label{eq:druwloss}
\end{equation}



\section{Experiments} \label{sec:exp}

We describe the dataset and preprocessing, as well as our experimental settings and results in this section.

\subsection{Dataset}

For all our experiments, we use the \ac{HVB} dataset as specified in the \ac{ACII} challenge.
The dataset contains over $59\,000$ instances for a total of over $36$\,hours of audio material, almost equally split into training, validation, and test sets.
The data comes from $1702$ speakers from four backgrounds: USA, China, South Africa, and Venezuela.
It contains seven different types of vocal bursts: cry, gasp, groan, grunt, laugh, pant, and scream -- with data that did not fall under any of those types being labelled as `other'.
Each instance has been annotated on a $[1-100]$ scale for each of the ten high-level emotional dimensions of: amusement, awe, awkwardness, distress, excitement, fear, horror, sadness, surprise, and triumph.
An average of $85.2$ raters have annotated each instance, and their individual ratings have been averaged to produce a gold standard annotation (which is then normalised to a $[0-1]$ range).
These gold standard annotations are used to derive arousal and valence values based on the circumplex model of affect~\cite{Russell80-CMA}.
The four tasks of the \ac{ACII} challenge are then defined as follows~\cite{Baird22-ACII}: Predict the ten high-level emotional dimensions for \avbhigh; Predict arousal and valence for \avbtwo; Predict a different high-level emotional dimension for each culture (a total of $40$ outputs) for \avbculture; Predict the type of vocal burst for \avbtype.

\begin{table*}[t]
    \centering
    \caption{Validation set results in terms of UAR for \avbtype, mean CCC, and mean $\rho$ for \avbtwo, \avbhigh and \avbculture, respectively. Shown are the best performing models for each task per architecture and loss weighting strategy, as well as the competition baseline score achieved with \etyou.}
    \begin{tabular}{l c c c c c c c}
    \toprule
    \textbf{Model} & \textbf{\avbtype} & \multicolumn{2}{c}{\textbf{\avbtwo}} &
    \multicolumn{2}{c}{\textbf{\avbhigh}} &
    \multicolumn{2}{c}{\textbf{\avbculture}} \\
    & UAR & CCC & $\rho$ & CCC & $\rho$ & CCC & $\rho$ \\
    \midrule
    baseline & .4166 & .4988 & - & .5638 & - & .4401 & - \\
    \hdashline
    \multicolumn{8}{c}{\textbf{Uniform Weighting}} \\
    $VANILLA$ & .5443 & .6964 & .6992 & .7205 & .7265 & .5892 & .5999 \\
    $CHAIN$ & .5534 & .6948 & .6979 & .7103 & .7180 & .5619 & .5844 \\
    $BRANCH$ & .5593 & .6934 & .6981 & .7114 & .7172 & .5791 & .5898 \\
    \hdashline
    \multicolumn{8}{c}{\textbf{Dynamic Weight Average}}\\
    $VANILLA$ & .5446 & .7026 & .7034 & .7271 & .7347 & .6025 & .6128 \\
    $CHAIN$ & \textbf{.5686} & \textbf{.7068} & \textbf{.7074} & \textbf{.7276} & \textbf{.7383} & 6070 & .6177 \\
    $BRANCH$ & .5479 & .6966 & .7008 & .7214 & .7272 & .5931 & 6043 \\
    \hdashline
    \multicolumn{8}{c}{\textbf{Restrained Revised Uncertainty Weighting}}\\
    $VANILLA$ & .5547 & .6992 & .7000 & .7213 & .7267 & .5892 & .5991 \\
    $CHAIN$ & .5588 & .6993 & .6989 & .7186 & .7237 & .5819 & .5961  \\
    $BRANCH$ & .5583 & .6955 & .6997 & .7123 & .7203 & .5850 & .5974  \\
    \hdashline
    \multicolumn{8}{c}{\textbf{Dynamic Restrained Uncertainty Weighting}}\\
    $VANILLA$ & .5447 & .6950 & .7002 & .7243 & .7341 & .6006 & .6130 \\
    $CHAIN$ & .5638 & .7019 & .7033 & .7291 & .7372 & \textbf{
    .6072} & \textbf{.6188}  \\
    $BRANCH$ & .5513 & .6951 & .6998 & .7204 & .7270 & .5917 & .6010 \\
    \bottomrule
    \end{tabular}
    \label{tab:validation_results}
\end{table*}

\subsection{Training}

We train our models for 30 epochs using a batch size of 8. AdamW \cite{Loshchilov2017} is chosen as the optimiser, with a learning rate of $10^{-5}$ for the feature extractor and $10^{-3}$ for the task networks. The rate is halved once the performance fails to improve for 5 epochs.
All audio clips are clipped or zero-padded to 2.5\,s length. For data augmentation, we use SpecAugment \cite{Park2019}, applying it to the input of the transformer part of \wav. Both time steps and features are masked with a probability of $0.05$.
We implement our models in PyTorch \footnote{https://pytorch.org/} and train them on Nvidia RTX3090 and A40 GPUs. Each model is run with multiple random initialisations.

\subsection{Results}

We report the performance in terms of UAR for \avbtype and the mean CCC and $\rho$ for each of \avbtwo, \avbhigh, \avbculture on the validation set. Our model architectures are designated \textit{VANILLA}, \textit{CHAIN}, and \textit{BRANCH}. Results are summarised in \cref{tab:validation_results}. We achieve a UAR of $.5686$ for \avbtype and a mean CCCs of $.7068$, $.7276$, and $.6072$ for \avbtwo, \avbhigh and \avbculture respectively. We also present the test set results of our best models in \cref{tab:test_results}.  

\begin{table}[]
    \centering
    \addtolength{\tabcolsep}{-5pt}
    \caption{Test set results for each task and best-performing model per architecture, as well as baseline results with \etyou.}
    \begin{tabular}{l c c c c}
    \toprule
    \textbf{Model} & \textbf{\avbtype} & \textbf{\avbtwo} & \textbf{\avbhigh} & \textbf{\avbculture} \\
    & UAR & CCC & CCC & CCC \\
    \midrule
    baseline & .4172 & .5084 & .5686 & .4401 \\
    \hdashline
    \vspace{0.1cm} \\
       $VANILLA$  & .5377 & .6938 & .7209 & .6020 \\
       $CHAIN$ & \textbf{.5618} & .6942 & .7261 & .6002 \\
    $BRANCH$ & .5418 & .6888 & .7148 & .5945 \\ 
    combined & .5560 & \textbf{.7066} & \textbf{.7363} & \textbf{.6195} \\
    \bottomrule
    \end{tabular}
    \label{tab:test_results}
\end{table}


\section{Discussion} \label{sec:discussion}

We base our discussion on the validation set results, since the test set is hidden and the number of evaluations limited.
All of our models surpass the baseline by a wide margin, demonstrating the efficacy of our approach using \wav as shared network for multi-task learning. Comparing the results shows that the simple \textit{VANILLA} architecture is competitive or even superior to the others for the emotions tasks, while being inferior for \avbtype. We interpret this as another indicator for the power of the \wav features. Using classifier chains led to the best overall results for all tasks, but in some cases degraded performance in later parts of the chain, indicating issues with error accumulation. Hyperparameter optimisation will likely improve the more complex attention architecture.
For the different loss weighting strategies, it can be seen that uniform weighting generally performs worst, which is expected since it is the least flexible approach and the weights were not tuned. The uncertainty-based strategies yield an improvement, with DRUW outperforming RRUW. Finally, DWA tends to perform best in the great majority of evaluations, demonstrating its efficacy. We note that for DRUW, further balancing of the contributions of DWA and RRUW may give even better results. 


\section{Conclusion} \label{sec:conclusion}

This paper presented a multi-task approach for predicting type and emotions of vocal bursts. Following previous work, we constructed models based on a large self-supervised feature extractor and investigated several architectures including classifier chains and task-specific multi-head attention networks. We combined these with loss balancing strategies based on uncertainty. Extensive experiments showed that our method far surpassed the shallow end-2-end baseline, demonstrating the effectiveness of self-supervised features and adaptive loss strategies for analysing vocal bursts. Future work could focus on optimal ordering of the classifier chains and experimentation with other self-supervised models.

\bibliography{lit.bib}

\begin{thebibliography}{10}
\providecommand{\url}[1]{#1}
\csname url@samestyle\endcsname
\providecommand{\newblock}{\relax}
\providecommand{\bibinfo}[2]{#2}
\providecommand{\BIBentrySTDinterwordspacing}{\spaceskip=0pt\relax}
\providecommand{\BIBentryALTinterwordstretchfactor}{4}
\providecommand{\BIBentryALTinterwordspacing}{\spaceskip=\fontdimen2\font plus
\BIBentryALTinterwordstretchfactor\fontdimen3\font minus
  \fontdimen4\font\relax}
\providecommand{\BIBforeignlanguage}[2]{{%
\expandafter\ifx\csname l@#1\endcsname\relax
\typeout{** WARNING: IEEEtran.bst: No hyphenation pattern has been}%
\typeout{** loaded for the language `#1'. Using the pattern for}%
\typeout{** the default language instead.}%
\else
\language=\csname l@#1\endcsname
\fi
#2}}
\providecommand{\BIBdecl}{\relax}
\BIBdecl

\bibitem{Hawk09-WORTH}
S.~T. Hawk, G.~A. Van~Kleef, A.~H. Fischer, and J.~Van Der~Schalk, ````worth a
  thousand words'': absolute and relative decoding of nonlinguistic affect
  vocalizations.'' \emph{Emotion}, vol.~9, no.~3, p. 293, 2009.

\bibitem{Sauter13-CRE}
D.~A. Sauter, C.~Panattoni, and F.~Happ{\'e}, ``Children's recognition of
  emotions from vocal cues,'' \emph{British Journal of Developmental
  Psychology}, vol.~31, no.~1, pp. 97--113, 2013.

\bibitem{Cordaro16-VCE}
D.~T. Cordaro, D.~Keltner, S.~Tshering, D.~Wangchuk, and L.~M. Flynn, ``The
  voice conveys emotion in ten globalized cultures and one remote village in
  bhutan,'' \emph{Emotion}, vol.~16, no.~1, p. 117, 2016.

\bibitem{Cowen19-BHV}
A.~S. Cowen, H.~A. Elfenbein, P.~Laukka, and D.~Keltner, ``Mapping 24 emotions
  conveyed by brief human vocalization,'' \emph{American Psychologist},
  vol.~74, no.~6, p. 698, 2019.

\bibitem{Huang19-NVS}
K.-Y. Huang, C.-H. Wu, Q.-B. Hong, M.-H. Su, and Y.-H. Chen, ``Speech emotion
  recognition using deep neural network considering verbal and nonverbal speech
  sounds,'' in \emph{Proc. ICASSP}.\hskip 1em plus 0.5em minus 0.4em\relax
  Brighton, UK: IEEE, 2019, pp. 5866--5870.

\bibitem{Hsu21-NVA}
J.-H. Hsu, M.-H. Su, C.-H. Wu, and Y.-H. Chen, ``Speech emotion recognition
  considering nonverbal vocalization in affective conversations,''
  \emph{IEEE/ACM Transactions on Audio, Speech, and Language Processing},
  vol.~29, pp. 1675--1686, 2021.

\bibitem{Baird22-EXVO}
A.~Baird, P.~Tzirakis, G.~Gidel, M.~Jiralerspong, E.~B. Muller, K.~Mathewson,
  B.~Schuller, E.~Cambria, D.~Keltner, and A.~Cowen, ``The {ICML} 2022
  expressive vocalizations workshop and competition: Recognizing, generating,
  and personalizing vocal bursts,'' \emph{arXiv preprint arXiv:2205.01780},
  2022.

\bibitem{Baird22-ACII}
A.~Baird, P.~Tzirakis, J.~A. Brooks, C.~B. Gregory, B.~Schuller, A.~Batliner,
  D.~Keltner, and A.~Cowen, ``The acii 2022 affective vocal bursts workshop \&
  competition: Understanding a critically understudied modality of emotional
  expression,'' \emph{arXiv preprint arXiv:2207.03572}, 2022.

\bibitem{Cowen2022HumeVB}
A.~Cowen, A.~Baird, P.~Tzirakis, M.~Opara, L.~Kim, J.~Brooks, and J.~Metrick,
  ``The hume vocal burst competition dataset {(H-VB)} | raw data [exvo: updated
  02.28.22] [data set],'' \emph{Zenodo}, 2022.

\bibitem{Pepino2021a}
L.~Pepino, P.~Riera, and L.~Ferrer, ``{Emotion Recognition from Speech Using
  wav2vec 2.0 Embeddings},'' \emph{Proc. Interspeech 2021}, pp. 3400--3404,
  2021.

\bibitem{Wagner22-Dawn}
J.~Wagner, A.~Triantafyllopoulos, H.~Wierstorf, M.~Schmitt, F.~Eyben, and B.~W.
  Schuller, ``Dawn of the transformer era in speech emotion recognition:
  closing the valence gap,'' \emph{arXiv preprint arXiv:2203.07378}, 2022.

\bibitem{Xin22-EES}
D.~Xin, S.~Takamichi, and H.~Saruwatari, ``Exploring the effectiveness of
  self-supervised learning and classifier chains in emotion recognition of
  nonverbal vocalizations,'' \emph{arXiv preprint arXiv:2206.10695}, 2022.

\bibitem{Jing22-RRT}
X.~Jing, M.~Song, A.~Triantafyllopoulos, Z.~Yang, and B.~W. Schuller,
  ``Redundancy reduction twins network: A training framework for multi-output
  emotion regression,'' \emph{arXiv preprint arXiv:2206.09142}, 2022.

\bibitem{Triantafyllopoulos22-ESR}
A.~Triantafyllopoulos, M.~Song, Z.~Yang, X.~Jing, and B.~W. Schuller,
  ``Exploring speaker enrolment for few-shot personalisation in emotional
  vocalisation prediction,'' \emph{arXiv preprint arXiv:2206.06680}, 2022.

\bibitem{Song22-DRU}
M.~Song, Z.~Yang, A.~Triantafyllopoulos, X.~Jing, V.~Karas, X.~Jiangjian,
  Z.~Zhang, Y.~Yoshiharu, and B.~W. Schuller, ``Dynamic restrained uncertainty
  weighting loss for multitask learning of vocal expression,'' \emph{arXiv
  preprint arXiv:2206.11049}, 2022.

\bibitem{Atmaja22-JPE}
B.~T. Atmaja, A.~Sasou \emph{et~al.}, ``Jointly predicting emotion, age, and
  country using pre-trained acoustic embedding,'' \emph{arXiv preprint
  arXiv:2207.10333}, 2022.

\bibitem{Read11-CC}
J.~Read, B.~Pfahringer, G.~Holmes, and E.~Frank, ``Classifier chains for
  multi-label classification,'' \emph{Machine learning}, vol.~85, no.~3, pp.
  333--359, 2011.

\bibitem{Baevski20-W2V}
A.~Baevski, Y.~Zhou, A.~Mohamed, and M.~Auli, ``wav2vec 2.0: A framework for
  self-supervised learning of speech representations,'' in \emph{Advances in
  Neural Information Processing Systems (NeurIPS)}, Vancouver, BC, Canada,
  2020, pp. 12\,449--12\,460.

\bibitem{Liu2019b}
S.~Liu, E.~Johns, and A.~J. Davison, ``End-to-end multi-task learning with
  attention,'' in \emph{Proceedings of the IEEE/CVF Conference on Computer
  Vision and Pattern Recognition (CVPR)}, June 2019.

\bibitem{Lin1989}
\BIBentryALTinterwordspacing
L.~I.-K. Lin, ``{A Concordance Correlation Coefficient to Evaluate
  Reproducibility},'' \emph{Biometrics}, vol.~45, no.~1, pp. 255--268, 1989.
  [Online]. Available: \url{http://www.jstor.org/stable/2532051}
\BIBentrySTDinterwordspacing

\bibitem{liu2021towards}
L.~Liu, Y.~Li, Z.~Kuang, J.~Xue, Y.~Chen, W.~Yang, Q.~Liao, and W.~Zhang,
  ``Towards impartial multi-task learning,'' in \emph{Proc. International
  Conference on Learning Representations}, Vienna, Austria, 2021.

\bibitem{liebel2018auxiliary}
L.~Liebel and M.~K{\"o}rner, ``Auxiliary tasks in multi-task learning,''
  \emph{arXiv preprint arXiv:1805.06334}, 2018.

\bibitem{Russell80-CMA}
J.~A. Russell, ``A circumplex model of affect,'' \emph{Journal of personality
  and social psychology}, vol.~39, no.~6, p. 1161, 1980.

\bibitem{Loshchilov2017}
\BIBentryALTinterwordspacing
I.~Loshchilov and F.~Hutter, ``Decoupled weight decay regularization,'' 2017.
  [Online]. Available: \url{https://arxiv.org/abs/1711.05101}
\BIBentrySTDinterwordspacing

\bibitem{Park2019}
D.~S. Park, W.~Chan, Y.~Zhang, C.-C. Chiu, B.~Zoph, E.~D. Cubuk, and Q.~V. Le,
  ``{SpecAugment: A Simple Data Augmentation Method for Automatic Speech
  Recognition},'' in \emph{Proceedings of Interspeech 2019}.\hskip 1em plus
  0.5em minus 0.4em\relax ISCA, Sep. 2019, pp. 2613--2617.

\end{thebibliography}
\bibliographystyle{IEEEtran}


\end{document}